# Flexible-to-semiflexible chain crossover on the pressure-area isotherm of lipid bilayer

I. N. Krivonos, S. I. Mukhin

Theoretical Physics Department, Moscow Institute for Steel and Alloys, Moscow, Russian Federation.

Corresponding author: S.I.Mukhin

e-mail: sergeimoscow@online.ru;   sergei@lorentz.leidenuniv.nl

*Abstract*

We found theoretically that competition between $\sim K_f q^4$ and $\sim Q q^2$ terms in the Fourier transformed conformational energy of a single lipid chain, in combination with inter-chain entropic repulsion in the hydrophobic part of the lipid (bi)layer, may cause a crossover on the bilayer pressure-area isotherm $P(A) \sim (A - A_0)^{-\alpha}$. The crossover manifests itself in the transition from α=5/3 to α=3. Our microscopic model represents a single lipid molecule as a worm-like chain with finite irreducible cross-section area $A_0$, flexural rigidity $K_f$ and stretching modulus $Q$ in a parabolic potential with self-consistent curvature $B(A)$ formed by entropic interactions between hydrocarbon chains in the lipid layer. The crossover area $A^*$ obeys relation $Q/\sqrt{K_f B(A^*)} \approx 2$. We predict a peculiar possibility to deduce effective elastic moduli $K_f$ and $Q$ of the individual hydrocarbon chain from the analysis of the isotherm possessing such crossover. Also calculated is crossover-related behavior of the area compressibility modulus $K_A$, equilibrium area per lipid $A_t$, and chain order parameter $S(\theta)$.

# 1. INTRODUCTION

Studying thermodynamics of the lipid bilayers that form biological membranes is of fundamental interest for understanding relation between membrane state and functioning of integral membrane proteins [1-3]. The latter are of vital importance for many processes in the living cells. Experimental data in lipid membranes indicate presence of a cross-over in the pressure-area isotherms $P(A) \sim (A - A_0)^{-\alpha}$ [4,5]. Formally, this means that exponent $\alpha$ changes substantially within some finite interval along the area axis $A$. A substantial amount of theoretical work has been devoted to description of the thermodynamic properties of lipid layers including pressure-area isotherms, chains order parameter as function of temperature, specific heat, etc. Theoretical approaches range from phenomenological Landau- de Gennes theory [6] to surface equations of states involving clustering [7-10] and raft formation [11]. Molecular dynamics [12] and Monte Carlo [13] simulations were done as well. Besides, the models were considered with phase transition due to change in the number of gauche conformations of the hydrocarbon chains [14-20], as well as models focused on the role of the excluded volume interactions between the chains [21,22].

These factors were also combined in the form of the additive area dependent contributions to the surface pressure [20].

In the previous work [23] a theoretical method has been proposed of calculation of the thermodynamic characteristics of lipid bilayer starting from a "microscopic" model of a smectic array of semi-flexible strings of finite length with a given flexural rigidity, see Fig. 1. The string is an idealized model of the hydrocarbon chain. The entropic repulsion between the neighboring chains in lipid membrane is modeled with effective potential. This entropic potential is then found self-consistently, by minimizing free energy of the bilayer, that, in turn, is calculated using path-integral integration over possible conformations of the strings. As a result, the lateral pressure profile inside the lipid bilayer was derived analytically, together with the area compressibility modulus and the temperature coefficient of area expansion of the membrane. In [23] only bending energy of the strings and entropic repulsion were included in the conformational energy functional. In the Fourier transformed representation the former energy is proportional to $\sim K_f q^4$, where $q$ is the wave-vector along the chain axis, and $K_f$ is chain flexural rigidity modulus. Then, resulting pressure-area isotherm of the lipid bilayer was derived in the form of a power law: $P_t(A) \sim (A - A_0)^{-\alpha}$, with constant exponent $\alpha = 5/3$. Here the lateral pressure of the lipid hydrocarbon chains (tails) $P_t(A)$ is expressed as a function of the area per lipid $A$ in the layer at a given temperature, with $A_0$ being the chain incompressible cross-section area. In the present work we added the stretching energy of the string to the energy functional [24]. In the Fourier transformed representation this energy is proportional to $\sim Qq^2$, where $Q$ is chain stretching modulus. Hence, our new chain energy functional contains now the sum:

$\sim K_f q^4 + Q q^2 + B(A)$, where $B(A)$ is self-consistently determined curvature of a parabolic effective entropic repulsive potential felt by a single chain due to surrounding chains in the lipid layer. The bending (flexural) energy dominates at large wave-vectors $q$, while the stretching energy dominates in the small $q$ limit. The entropic repulsion term $B(A)$ sets an upper limit for the wave-vectors $q$ that are essential for thermodynamics. The entropic repulsion increases with a decrease of area per lipid in the layer, i.e. parameter $B(A)$ becomes greater when $A \to A_0$, making big $q's$ important.

As a result, the bending energy term $\sim K_f q^4$ dominates, and we recover the previous pressure-area isotherm [23] with the exponent $\alpha = 5/3$. On the other hand, when the area per lipid increases, the entropic repulsion becomes weaker, and parameter $B(A)$ becomes smaller. Hence, important $q's$ become smaller too and stretching energy term $\sim Q q^2$ dominates. As a result, a new exponent arises [24]: $\alpha = 3$, corresponding to the stretching dominated conformational energy of the strings, Figs. 2, 3. We found, that the crossover between the two area per lipid regions with different values of the exponent $\alpha$ takes place at the area per lipid $A^*$ determined from the condition: $Q/\sqrt{K_f B(A^*)} \approx 2$. The physical states of the lipid layer in the two regions separated by crossover differ by a substantial change of the value of the chain order parameter, that characterizes deviation of the chain from the straight line, Fig. 7.

Also calculated are elastic moduli of the membrane and their dependences on temperature and on the microscopic elastic moduli of the individual chains constituting the lipid bilayer Figs. 4-6. Finally, we discuss how fitting of experimental isotherms of lipid bilayers with our theoretical isotherms may help to deduce elastic moduli $K_f$ and $Q$ of individual lipid chains constituting the membrane.

The plan of the article is as follows. In the first part we formulate the physical model [24] of bilayer and review the path-integral method of summation over all conformations of idealized hydrocarbon chain [23]. In the second part we derive and solve analytically (in two different limits) a self-consistency equation for the curvature $B(A)$ of the effective parabolic entropic potential in the layer. The pressure-area isotherms are then derived in the analytical form. In the third through fifth parts of the paper we present results of calculations of the thermodynamic and elastic characteristics of the whole bilayer that follow from our microscopic model. In the Conclusions we discuss correspondence of the theoretical results with available experimental data, as well as consider applicability region of the used approximations.

## 2. ENERGY FUNCTIONAL OF LIPID LAYER: STRING MODEL

Hydrocarbon chain (see Fig.1) is modeled as a flexible string with flexural rigidity $K_f$ and stretching modulus $Q$, while entropic repulsion with surrounding lipid chains is modeled via parabolic potential with self-consistently determined curvature $B (= B(A))$:

$$E = \int_0^L \left( K_f \left( \frac{d^2\mathbf{R}}{dz^2} \right)^2 + Q \left( \frac{d\mathbf{R}}{dz} \right)^2 + B\mathbf{R}^2 \right) dz \qquad (1)$$

where $L$ is the chain length. Here we consider only small deviations of the chains from the straight line: $|\mathbf{R}(z)| \ll L$. The latter condition is fulfilled when: $\sqrt{<\mathbf{R}^2(z)>}/L \leq \left(k_B T/L^2 P_{eff}\right)^{1/2} \ll 1$, where effective tension in the bilayer is $P_{eff}$, and $k_B$ is the Boltzmann constant. The returning force $-B\mathbf{R}(z)$ acts against deviation $\mathbf{R}(z)$ of the chain from the vertical straight line, where coordinate $z$ measures depth inside the lipid layer with the hydrophilic (polar) heads residing at the layer surface $z = 0$, while hydrophobic (non-polar) tails formed of hydrocarbon chains constitute the body of the slab $0 < z \leq L$. Here $\mathbf{R}(z)$ is a vector in $\{x, y\}$ plane, characterizing deviation from the $z$-axis: $\mathbf{R}^2(z) = X^2(z) + Y^2(z)$.

Hence, the energy of a single string consists of the three parts: bending, stretching and effective entropic potential energies. In an equilibrium the effect of the entropic repulsion is compensated by the inter-chain attraction due to van der Waals interactions and by surface tension energy $\gamma \cdot A$, where $\gamma$ is

surface tension at hydrophobic-hydrophilic interface [4]:

$$P_t(A(T)) = P_{eff} = \gamma + P_{hg} + P_{vdW}.$$

Here the total tension in the bilayer is zero. The integral $P_t = \int_0^L \Pi_t(z)dz$ of the repulsive (chain) part of the lateral pressure profile $\Pi_t(z)$ over the bilayer hydrophobic region equals the balancing effective tension in the bilayer $P_{eff}$, that besides the surface tension $\gamma$, includes head group repulsion of electrostatic origin $P_{hg}$, pressure arising from van der Waals interactions between chains $P_{vdW}$, etc.. Here $A(T)$ is equilibrium area per lipid at a given temperature. We choose $P_{eff} = 100 \text{dyn/cm} > \gamma \sim 50 \text{dyn/cm}$, because attractive dispersion interactions between hydrocarbon chains are included in the effective surface tension [29]. In general, at room temperature, effective surface tension for typical lipid bilayer belongs to the interval: $50 \leq P_{eff} \leq 150$ dyn/cm [2],[29]. Considering for simplicity single chain consisting of $N$ links of equal length $a$, we rewrite the Fourier transformed energy functional (1):

$$E = L^2 \int_{-\pi/a}^{\pi/a} |R_q|^2 \left(K_f q^4 + Q q^2 + B\right) \frac{dq}{2\pi} = L \sum_q |R_q|^2 E_q, \qquad (2)$$

where summation in wave-vector $q$ is made over the interval $2\pi/a$ with the step $2\pi/L$; $E_q = K_f q^4 + Q q^2 + B$; and $R_q$ is Fourier image of the function $R(z)$. Since we consider membrane that is isotropic in the

{x, y}-plane, the x- and y-components of the vector field $R(z)$ give equal contributions into partition function of the string, and we obtain:

$$Z = Z_x Z_y = Z_x^2 = \left( \prod_q \int_0^\infty \exp\left( -\frac{LE_q}{k_B T} |X_q|^2 \right) |X_q| d|X_q| \right)^2 = \left( \prod_q \frac{k_B T}{2LE_q} \right)^2 \qquad (3)$$

Then the free energy $F = -k_B T \ln Z$ ($k_B$ is the Boltzmann constant) can be expressed as follows:

$$\Delta F = F(B) - F(B=0) = 2k_B T \sum_q \ln\left( \frac{E_q}{E_q(B=0)} \right) \qquad (4)$$

where free energy of a free single chain is subtracted for convergence of the sum.

## 3. PRESSURE-AREA ISOTHERM OF LIPID LAYER EXPRESSED VIA CURVATURE OF THE INTER-CHAIN ENTROPIC POTENTIAL

The self-consistent equation for the curvature $B$ of the parabolic potential modeling inter-chain entropic repulsion in the lipid layer takes the form:

$$\frac{\partial F}{\partial B} = 2k_B T \sum_q \frac{1}{E_q} = <R^2> L = \frac{\left(\sqrt{A} - \sqrt{A_0}\right)^2}{\pi} L \qquad (5)$$

where $A_0$ is incompressible area of the chain cross-section: $A_0 = V_0/L$, $V_0$ is incompressible volume of the lipid chain; $\left(\sqrt{A} - \sqrt{A_0}\right)^2$ is area swept by the string formed with the centers of the chain cross-sections.

Obviously, when area per lipid $A$ is close to the incompressible area $A_0$ the self-consistent curvature $B$ is big, and hence we may pass from summation to the integration over the wave-vector $q$ in Eq. (5):

$$\left(\sqrt{A} - \sqrt{A_0}\right)^2 L = k_B T L \int_{-\pi/a}^{\pi/a} \frac{dq}{E_q} \qquad (6)$$

The resulting self-consistency equation for the curvature $B$ is:

$$\left(\sqrt{A} - \sqrt{A_0}\right)^2 = \frac{\pi}{\sqrt{2}} \frac{k_B T}{K_f^{1/4} B^{3/4}} \left(1 + Q/\left(2\sqrt{BK_f}\right)\right)^{-1/2} \qquad (7)$$

This equation is solved numerically, but analytical results are also obtained in the two limiting cases: $\xi \ll 1$ and $\xi \gg 1$, where $\xi \equiv Q/(2\sqrt{BK_f})$. These two limits correspond to the domination of the bending energy and of the stretching energy respectively:

$$B(A) = \frac{1}{K_f^{1/3}} \frac{\pi^{4/3}(k_B T)^{4/3}}{\sqrt[3]{4}} \left(\sqrt{A} - \sqrt{A_0}\right)^{-8/3}, \quad \xi \ll 1; \qquad (8)$$

$$B(A) = \frac{\pi^2 (k_B T)^2}{Q} \left(\sqrt{A} - \sqrt{A_0}\right)^{-4}, \quad \xi \gg 1. \qquad (9)$$

Derived here relation (8) coincides with the previous result expressed in Eq. (A5) in [23], that was obtained by a more detailed method without use of the Fourier transformed version of the energy functional (2) and with the stretching energy not included.

As long as $\xi = \xi(A)$, due to dependence on area $A$ of the curvature $B = B(A)$, the crossover from the limit (8) to the limit (9) may take place at some particular area per lipid $A*$ determined from the condition: $\xi(A*) \approx 1$. This crossover area $A*$ is found below (see Eq. (14)).

Substituting self-consistent solution $B = B(A)$ into the chain free energy (4) one finds the lateral pressure - area per lipid isotherm for the hydrophobic part of the lipid (bi)layer:

$$P_t \equiv -\frac{\partial \Delta F(A)}{\partial A} = \frac{\partial B}{\partial A} \sum_q \frac{2k_B T}{E_q} \qquad (10)$$

Since the sum on the right hand side in Eq. (10) enters also the self-consistency equation (5), the pressure-area isotherm can be expressed in the closed analytical form:

$$P_t(A) = \pi^{-1} \left(\sqrt{A} - \sqrt{A_0}\right)^2 L \frac{\partial B(A)}{\partial A} \qquad (11)$$

with the curvature $B(A)$ directly involved.

## 4. CROSSOVER ON THE ISOTHERM OF LIPID LAYER: COMMUNICATION BETWEEN "MACRO" AND "MICRO"

It is remarkable that crossover in the isotherm of a lipid (bi)layer provides, in principle, an intriguing possibility to follow link between macro- and microscopic properties of the biomembrane. Namely, one is able to deduce effective elastic moduli of the individual lipid chains constituting the bilayer from the pressure-area isotherm of the whole macroscopic system.

In order to explain the deduction "recipe" (see Eqs. (15) (16) below), we consider theoretical predictions following from the general model described in the previous section. First, we substitute successively equations (8) and (9) into general equation (11) and find (in the limit of small enough areas per lipid: $A/A_0 \cong 1$) the following analytical formulas describing the lateral pressure – area isotherms of the lipid layer:

$$\frac{P_t}{k_B T} = \frac{2}{3}\left(\frac{\pi k_B T}{4 K_f}\right)^{1/3} V A^{-3/2} \left(\sqrt{A} - \sqrt{A_0}\right)^{-5/3}, \quad \xi << 1; \qquad (12)$$

$$\frac{P_t}{k_B T} = \frac{2\pi k_B T}{Q} V A^{-3/2} \left(\sqrt{A} - \sqrt{A_0}\right)^{-3}, \quad \xi >> 1, \qquad (13)$$

where $V = AZ_m \approx AL$ is (conserved) volume per lipid molecule in the hydrophobic part of the lipid layer, $Z_m$ is actual thickness of the hydrophobic part of the lipid layer, and $Z_m \approx L$ in the limit of small deviations from the straight line of the string modeling the chain. The asymptotic relation (12) valid for the bending dominated free energy is the same as in previous work [23], while relation (13) is new and corresponds to domination of the stretching energy. The crossover region between these two limiting cases is difficult to express analytically, but result of numerical calculation using equations (7) and (11) is presented in Fig. 2 together with the two curves corresponding to the analytical results given in (12) and (13). Before considering Figs. 2, and 3 it is worth to find the crossover area per lipid $A*$ using definition $\xi(A*) \equiv Q/2\sqrt{B(A*)K_f} \approx 1$ and substituting asymptotic equations (8) and (9) for $B(A*)$, that both lead to the following result:

$$A^* = \left(\sqrt{A_0} + \sqrt{2\pi \cdot kT \sqrt{K}}/Q^{3/4}\right)^2 \quad (14)$$

It follows from (14) that at $Q \to 0$ we have $A^* \to \infty$ and, hence, the crossover to stretching dominated region of areas per lipid is shifted away from the interval of reasonable areas $A$ : $A \geq A_0$. Therefore, for too small stretching modulus $Q$ the bending dominated dependence *P(A)* derived in (8) occupies the whole *A*-axis and cross-over is not encountered. It is convenient to evaluate relative strength of the stretching and bending energies of a chain by a dimensionless parameter $\sigma = QL^2/K_f$. Hence, bending dominated isotherm is marked with the dimensionless parameter $\sigma = QL^2/K \to 0$ in Fig. 2. In the opposite limit $K \to 0$ it follows from (14) : $A^* \to A_0$, i.e. the whole *A*-axis is occupied by the stretching dominated region, and the corresponding stretching dominated isotherm derived in (9) is marked with $\sigma \to \infty$ label in Fig. 2.

The third isotherm in Fig. 2, labeled $\sigma \cong 10^2$, exhibits the crossover from the bending dominated region at small areas per lipid, $A \leq A^*$, to the stretching dominated interval at greater areas, $A \geq A^*$. As it follows from the Eq. (14), the bending dominated region $A_0 \leq A \leq A^*$ shrinks when temperature *T* decreases.

It proves to be that more informative than isotherms themselves are the plots of their logarithmic derivatives presented in Fig.3, i.e.: $d\ln(P_t)/d\ln(A - A_0) \approx -\alpha(A)$ vs $A/A_0$. This becomes obvious after writing the isotherm equations (12) and (13) in the limit $A \to A_0$:

$$P_t \approx \frac{4\pi^{1/3}(k_B T)^{4/3} LA_0^{1/3}}{3K_f^{1/3}} (A-A_0)^{-5/3}, \alpha = 5/3, \xi \ll 1; \quad (15)$$

$$P_t \approx \frac{16\pi(k_B T)^2 LA_0}{Q} (A-A_0)^{-3}, \alpha = 3, \xi \gg 1. \quad (16)$$

Hence, fitting experimental isotherm of lipid bilayer with the hyperbolas from Eq. (15) or Eq. (16) it is possible, in principle, to determine which case: $\xi \ll 1$ or $\xi \gg 1$, corresponds to the state of the lipid bilayer . After that one can deduce relevant effective "microscopic" elastic moduli $K_f$ and $Q$ of *an individual lipid chain* using Eqs. (15) and (16) , that contain these parameters as pre-factors.

## 5. CHAIN ORDER PARAMETER AND MACROSCOPIC ELASTIC MODULI OF LIPID BILAYER FROM THE STRING MODEL

To make numerical estimates based on our model of lipid (bi)layer , Eq. (1), we use the following parameters: chain length $L=15$ Å, chain incompressible area

$A_0=20$ Å$^2$, and take $T_0=300$ K as a reference temperature. The chain flexural rigidity is defined as [25]: $K_f=EI$, where $E \approx 0.6 GPa$ - is chain Young modulus [26], $I = A_0^2/4\pi$ - is (geometric) moment of inertia. The flexural rigidity can also be evaluated from the polymer theory [27]: $K_f = k_B T l_p$, where $l_p \approx L/3$ is chain persistence length [26]. Both estimates give approximately $K_f \approx k_B TL/3$ at chosen $L$ and at $T=T_0$. An estimate of the value of stretching modulus $Q$ can be done using energy functional (1) and assuming that both contributions due to bending and stretching energies are of the same order, leading to: $Q \sim 10^{-6} \div 10^{-5} dyn$. Eq. (17) is valid in the area interval with dominant bending energy, while Eq. (18) applies for the area interval with dominant stretching energy respectively. When areas per lipid

Differentiation of $P_t(A)$ gives the area compressibility modulus:

$K_a = -AdP_t(A,T)/dA$ as function of area per chain and temperature. Analytical expressions for this modulus are then derived using Eqs. (12) and (13):

$$K_a = VkT\left(\frac{\pi k_B T}{4K_f}\right)^{1/3}\left(A^{-3/2}\left(\sqrt{A}-\sqrt{A_0}\right)^{-5/3} + \frac{5}{9}A^{-1}\left(\sqrt{A}-\sqrt{A_0}\right)^{-8/3}\right), \xi \ll 1; \quad (17)$$

$$K_a = \frac{3\pi V(kT)^2}{Q}\left(A^{-3/2}\left(\sqrt{A}-\sqrt{A_0}\right)^{-3} + A^{-1}\left(\sqrt{A}-\sqrt{A_0}\right)^{-4}\right), \xi \gg 1. \quad (18)$$

Eq. (17) is valid in the area interval with dominant bending energy, while Eq. (18) applies for the area interval with dominant stretching energy respectively. When areas per lipid are close enough to $A_0$ expressions in (17) and (18) can be further simplified by retaining only most diverging terms. In this way we conclude that under a decrease of area per lipid $A$ (shrinking of the bilayer) the dependence of the area compressibility modulus on area per lipid $K_a(A)$ should change from $K_a \sim (A - A_0)^{-4}$ in the stretching dominated region to $K_a \sim (A - A_0)^{-8/3}$ in the bending dominated region. Calculated absolute value of the modulus $K_a$ using (18) with other parameters: $T = 300\ K$, $L = 15\ Å$, $A_0 = 20\ Å^2$, $P_{eff} = 100$ dyn/cm and $Q \approx 10^{-6} dyn$, is $K_a \sim 420\ erg/cm^2$. This theoretical value is well comparable with the known data $K_a \approx 300\ erg/cm^2$ [29]. Calculated temperature dependence of the equilibrium area $A_t$ for a fixed value of $\sigma = QL^2/K_f \sim 10^2$ is shown in Fig. 4. From this data we find also the temperature coefficient of area expansion: $K_T = dA_t/A_t dT \cong 0.9 \cdot 10^{-3} K^{-1}$, in good correspondence with available data [28], [29]: $K_T \sim 10^{-3} K^{-1}$.

It is interesting also to find how the area compressibility modulus depends on the relative strength of the stretching energy with respect to the bending energy of the lipid chain, which is reflected by the dimensionless parameter $\sigma = QL^2/K_f$ introduced earlier. In Fig. 5 our theoretical results are presented, that demonstrate an increase of the area compressibility modulus $K_a$ of the lipid bilayer as function of the parameter $\sigma$. Though, an increase of $K_a$ by approximately three times necessitates corresponding increase of parameter $\sigma$ by 3 orders of magnitude. Simultaneously, under such an increase of parameter $\sigma$ the equilibrium area per lipid $A_t$ in the layer has decreased approximately by a factor of *2*, as it follows from our results presented in Fig. 6. The curves in Figs. 5, 6 together may be interpreted as lateral "hardening" of the lipid layer due to shrinking of the average nearest-neighboring inter-chain distances when stretching energy is added. This is understandable if one realizes that the stretching energy makes wiggling of the lipid tails energetically unfavorable. The latter, in turn, "pushes" the lipid layer more close to a gel-like state with higher area density of lipids. This semi-intuitive explanation is further supported by our calculations of the chain order parameter $S(\theta) = \frac{1}{2}(3<\cos^2(\theta)> - 1)$, where the local tangent angle $\theta(z)$ is averaged along the length of the chain $0 \leq z \leq L$, and the thermodynamic average over different chain conformations is also performed. We use for calculations the approximate relation valid for small deviations of the chain from the vertical straight line (parallel to z-axis in Fig. 1):

$$1 - <\cos^2\theta> \approx <tg^2(\theta)> \propto <(X'(z))^2> = \frac{1}{2\pi}\int_{\pi/a}^{\pi/a} \frac{q^2 T}{E_q} dq \quad (19)$$

Since the order parameter belongs to an interval $0 \leq S(\theta) \leq 1$, approximate relation in Eq. (19) is valid as long as deviations of the chains from the straight line are small: $\theta \sim \frac{\sqrt{<X^2(z)>}}{L} << 1$. Thus, besides being important characteristic of the prevailing conformations of the chains in the lipid bilayer, the order parameter provides a consistency check. The results of calculations are presented in Fig. 7. It is easier to understand the origin of the different limiting curves in Fig. 7 when rewriting Eq. (19) using definition given after Eq. (3):

$$E_q = K_f q^4 + Qq^2 + B \quad (20)$$

Then, in the limit $\sigma \to 0$, when bending energy dominates, the term proportional to $q^2$ in $E_q$ vanishes. Simultaneously, due to the bending energy contribution ($\sim K_f q^4$), the integral in Eq. (19) converges at small wave-vectors $q \leq (B(A)/K_f)^{1/4}$, leading to a decrease of the order parameter with increasing area per lipid: $S(\theta) \approx 1 - const \times (B(A))^{-1/4}$, where: $B(A) \sim (A - A_0)^{-8/3}$, according to Eq. (8). In the opposite limit $\sigma \to \infty$, when stretching energy dominates, we obtain for the integral in Eq. (19) the following expression:

$$< (X'(z))^2 > = \frac{1}{2\pi} \int_{-\pi/a}^{\pi/a} \frac{q^2 dq}{Qq^2 + B(A)} = \frac{T}{aQ}\left(1 - \sqrt{\frac{B}{Q}} \frac{L}{N\pi} \cdot arctg\left(\sqrt{\frac{Q}{B}} \frac{N\pi}{L}\right)\right) \qquad (21)$$

Hence, in the limit: $\sqrt{B(A)/Q}\,(L/N\pi) >> 1$, i.e. when $A \to A_0$, we find $< (X')^2 > \to 0$, and therefore: $S(\theta)_{\sigma \to \infty} \to 1$. In the opposite limit: $\sqrt{B(A)/Q}\,(L/N\pi) << 1$, i.e. when $A$ sufficiently exceeds $A_0$, $< (X')^2 > \approx NT/LQ$, and, hence, the order parameter is practically area-independent: $S(\theta)_{\sigma \to \infty} \cong const < 1$, as is seen in Fig. 7, where $S(\theta)_{\sigma \to \infty} \equiv S_Q$ becomes flat as $A/A_0$ increases. In the intermediate case, see curve labeled $\sigma = 10^2$, the order parameter as function of the area per lipid interpolates qualitatively between the two limiting dependences just described.

Finally, we observe, that as it follows from Fig. 7, our approximation of small deviations of the chain from a straight line is valid under chosen numerical values of the basic parameters of the lipids up to areas per lipid $A/A_0 \leq 3$, since deviation of the order parameter $S(\theta)$ from 1 in all considered regimes proves to be small: $1 - S(\theta) \leq 0.3$.

# 6. CONCLUSIONS

We derived analytical expressions for the pressure-area isotherms of a lipid bilayer using string model of hydrocarbon chains, that includes flexural and stretching moduli of a single chain, as well as self-consistent entropic repulsion acting between the chains in the lipid bilayer. A crossover on the pressure – area isotherm is predicted, that arises due to competition between bending and stretching contributions to the total conformational energy of the individual chains. A theoretical method of the data analysis is proposed that, in principle, permits to deduce microscopic effective elastic moduli of the individual lipid molecules by studying pressure-area isotherms of the macroscopic lipid (bi)layer.

The applicability criteria and checks of the theory using comparison with known experimental and numerical simulation data for lipid bilayers are presented. A generalization of the proposed model for a description of the spatially inhomogeneous thermodynamic states of lipid bilayers is in progress now.

# 7. REFERENCES

1. R.S. Cantor, Boiphys. J. **76**, 2625 (1999).
2. D. Marsh, Biochim. Biophys. Acta **1286**, 183 (1996).
3. S.I. Sukharev, W.J. Sigurdson, C. Kung et al., J. Gen. Physiol. **113**, 525 (1999).
4. A. Ben-Shaul, Structure and Dynamics of Membranes, Elsevier Since, Amsterdam, 1995.
5. A.F. Mingotaud, C. Mingotaud, L.K. Patterson, Handbook of Monolayers, Academic, San Diego (1993), Vols 1 and 2.
6. Ping Sheng, Phys. Rev. Lett. **37**, 1059 (1976).
7. J.N. Israelachvili, Langmuir **10**, 3774 (1994)
8. E. Ruckenstein, B. Li, Langmuir **12**, N 9, 2308-2315 (1996)
9. E. Ruckenstein, B. Li, J. Phys. Chem. **102**, 981-989 (1998)
10. P.I. Kuzmin, S.A. Akimov, Yu.A. Chizmadzhev, J. Zimmerberg, F.S. Cohen, Biophys. J. **88**, 1120 (2005).
11. J.L. Harden, F.C. MacKintosh, P.D. Olmsted, Phys. Rev. E **72,** 011903 (2005).
12. A.J. Kox, J.P.J. Michels, F.W. Wiegel, Nature (London) **287**, 317 (1980).
13. A. Georgallas, D.A. Pink, J. Colloid Interface Sci. **62**, 125 (1977).
14. S. Marcelja, Biochim. Biophys. Acta **367**, 165 (1974).
15. J.F. Nagle, J. Membrane Biol. **27**, 233 (1976).

16. A. Caille, D.A. Pink, F. De Verteuil, M.J. Zuckermann, Can. J. Phys. **58**, 581 (1980).
17. S. Doniach, J. Chem. Phys. **68**, 4912 (1978).
18. A. Caille, A. Rapini, M.J. Zuckermann, A. Cros, S. Doniach, Can. J. Phys. **56**, 348 (1978).
19. F. Jahnig, J. Chem. Phys. **70**, 3279 (1979).
20. R.S. Cantor and K.A. Dill, Langmuir **2**, 231 (1986).
21. J.-L. Firpo, J.J. Dupin, G. Albinet, A. Bois, *et. al.*, J. Chem. Phys. **68**, 1369 (1978).
22. J.J. Dupin, J.-L. Firpo, G. Albinet, A. Bois, *et. al.*, J. Chem. Phys. **70**, 2357 (1979).
23. S.I. Mukhin, S. Baoukina. , Phys. Rev. **E 71**, 061918 (2005).
24. I.N. Krivonos and S.I. Mukhin, Biophys. J. 92 (1) Part 2 Suppl., 583a (2007).
25. L.D. Landau and E.M. Lifshitz, *Mechanics* (Pergamon Press, 1960).
26. *Polymer encyclopedia* **V. 1-3** (BSE, 1977).
27. D. Nelson. *Defects and Geometry in Condensed Matter Physics* (Cambridge, 2002).
28. R. Waugh, E.A. Evans, Biophys. J. **26**, 115-132 (1979).
29. E. Lindahl, and O. Edholm, J. Chem. Phys. **113**, 3882 (2000).

# 8. FIGURE CAPTIONS

Fig. 1. Model of lipid membrane in mean-field approximation. Left panel: sketch of lipid monolayer (the mirror-symmetric part of a bilayer is not shown). Hydrocarbon tails of lipid molecules form hydrophobic part of lipid monolayer. Hydrophilic polar heads (filled ellipses) form hydrophobic-hydrophilic interface. Right panel: mean-field flexible string model of the hydrophobic layer. The arrows indicate entropic forces $\pm BX$, acting on a hydrocarbon chain in the self-consistent entropic potential $BX^2$, that arises due to surrounding neighbors.

Fig. 2. Calculated pressure – area per lipid isotherms for the lateral pressure $P_t$ in the hydrophobic (tails) part of the lipid bilayer with different relative strengths of the single chain stretching and bending energies characterized by dimensionless parameter $\sigma = QL^2/K_f$ (see text). The isotherm labeled with $\sigma \to 0$ corresponds to dominant bending energy of the chains (see text Eq. (12)); the curve labeled with $\sigma \to \infty$ corresponds to dominant stretching energy of the chains (see text Eq. (13)). The curve labeled $\sigma \approx 10^2$ corresponds to an intermediate case and possesses a crossover between the two isotherms drawn in the two limits mentioned above. $A_0$ is incompressible area of the hydrocarbon chain. The temperature for all the curves is $T = 300K$.

Fig. 3. The same cases as in Fig. 2, but for calculated logarithmic derivative of lateral pressure with respect to area per lipid (subtracted is incompressible area of hydrocarbon chain $A_0$). The logarithmic derivative gives value of the exponent $-\alpha(A)$ in the limit $A \to A_0$ in the isotherm equation of state of the lipid bilayer: $P_t \sim (A - A_0)^{-\alpha(T)}$. The middle curve ($\sigma \approx 10^2$) demonstrates crossover between exponents $\alpha = 5/3$ and $\alpha = 3$ that are marked with the straight dashed lines. The two other curves are exact dependences given in Eqs. (12) and (13) describing the isotherms in the limits of dominant bending ($\sigma \to 0$) and stretching ($\sigma \to \infty$) energy of the hydrocarbon chains respectively. The temperature for all the curves is $T = 300K$.

Fig. 4. Calculated temperature dependence of the equilibrium area per lipid $A_t$ in the lipid bilayer for a fixed value of $\sigma \sim 10^2$ characterizing ratio of the chain stretching and bending energies (see text). Other parameters are: $T_0 = 300K$, $L=15$ Å, $A_0=20$ Å$^2$, $P_{eff} = 100$ dyn/cm, where $L$ is chain length, $P_{eff}$ is total lateral pressure in the hydrophobic part of the lipid bilayer, $A_0$ is incompressible area of the hydrophobic chain.

Fig. 5. Calculated area compressibility modulus $K_a$ of the lipid bilayer as function of the dimensionless parameter $\sigma = QL^2 / K_f$. Other parameters are as in Fig. 4. The greater values of parameter $\sigma$ correspond to increasing relative strength of the stretching energy of a hydrocarbon chain with respect to its bending energy.

Fig. 6. Calculated equilibrium area per lipid $A_t$ in the lipid bilayer as function of dimensionless parameter $\sigma = QL^2 / K_f$ characterizing relative strength of the stretching and bending energies of the individual hydrocarbon chains in the lipid bilayer. The greater values of parameter $\sigma$ correspond to increasing relative strength of the stretching energy of a hydrocarbon chain with respect to its bending energy. Other parameters are as in Fig.4.

Fig.7. Calculated chain order parameter $S(\theta) = \frac{1}{2}(3<\cos^2(\theta)> - 1)$ as function of area per lipid in a lipid (bi)layer. $A_0$ is incompressible area of hydrocarbon chain. The local tangent angle $\theta(z)$ is averaged along the length of the chain $0 \leq z \leq L$, as well as over different chain conformations. The curve labeled with $\sigma \to 0$ corresponds to dominant bending energy of the chains; the curve labeled with $\sigma \to \infty$ corresponds to dominant stretching energy of the chains; the curve labeled $\sigma \approx 10^2$ corresponds to an intermediate case and possesses a crossover between the curves drawn in the two limits mentioned above. Other parameters are as in Fig. 4.

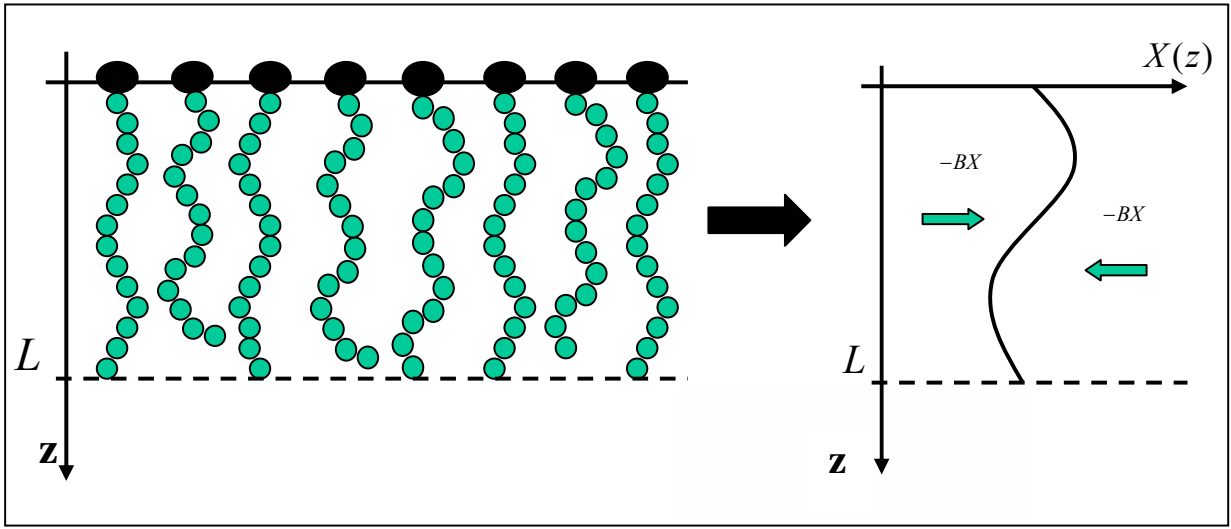

Fig. 1

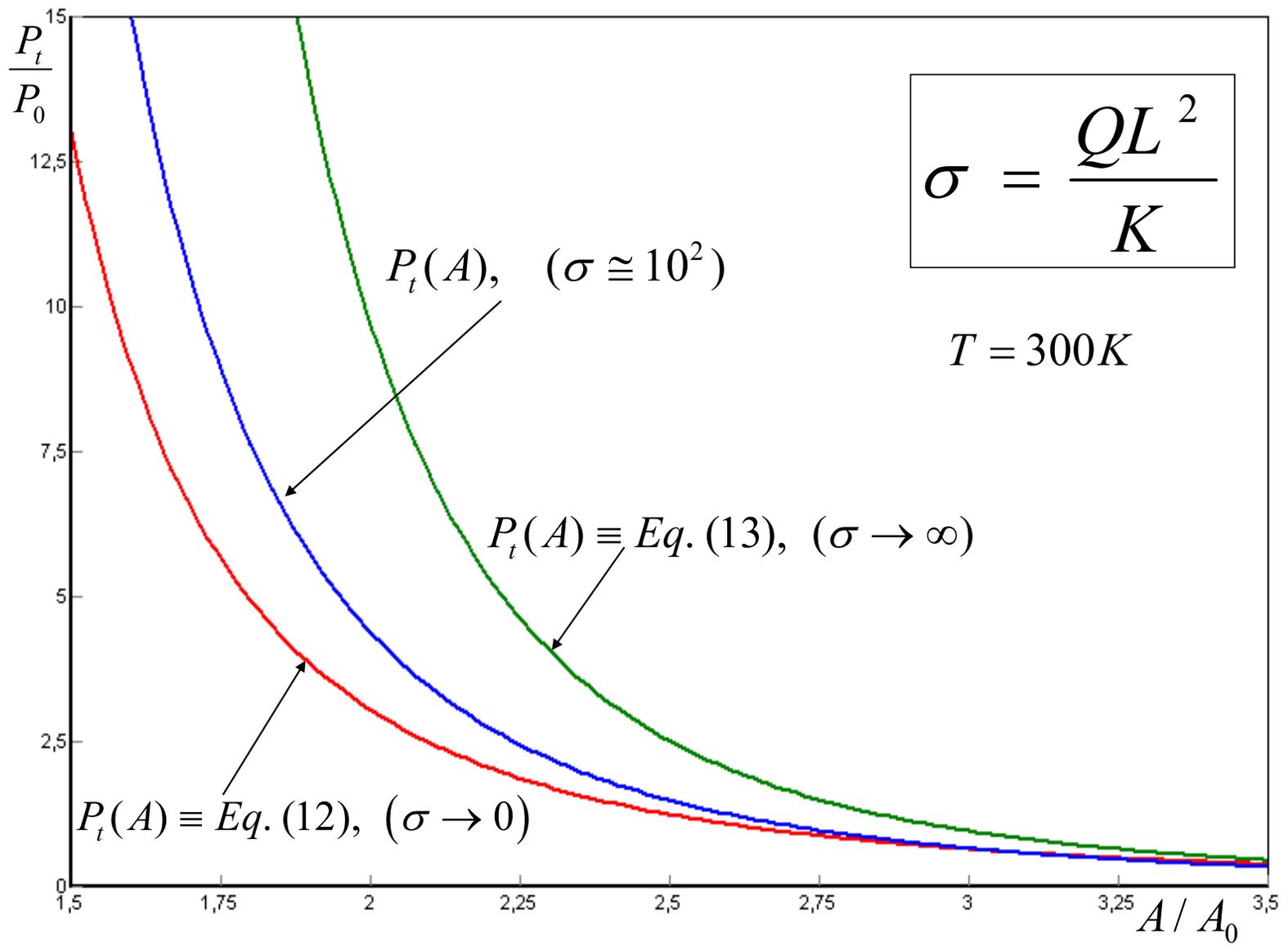

Fig. 2

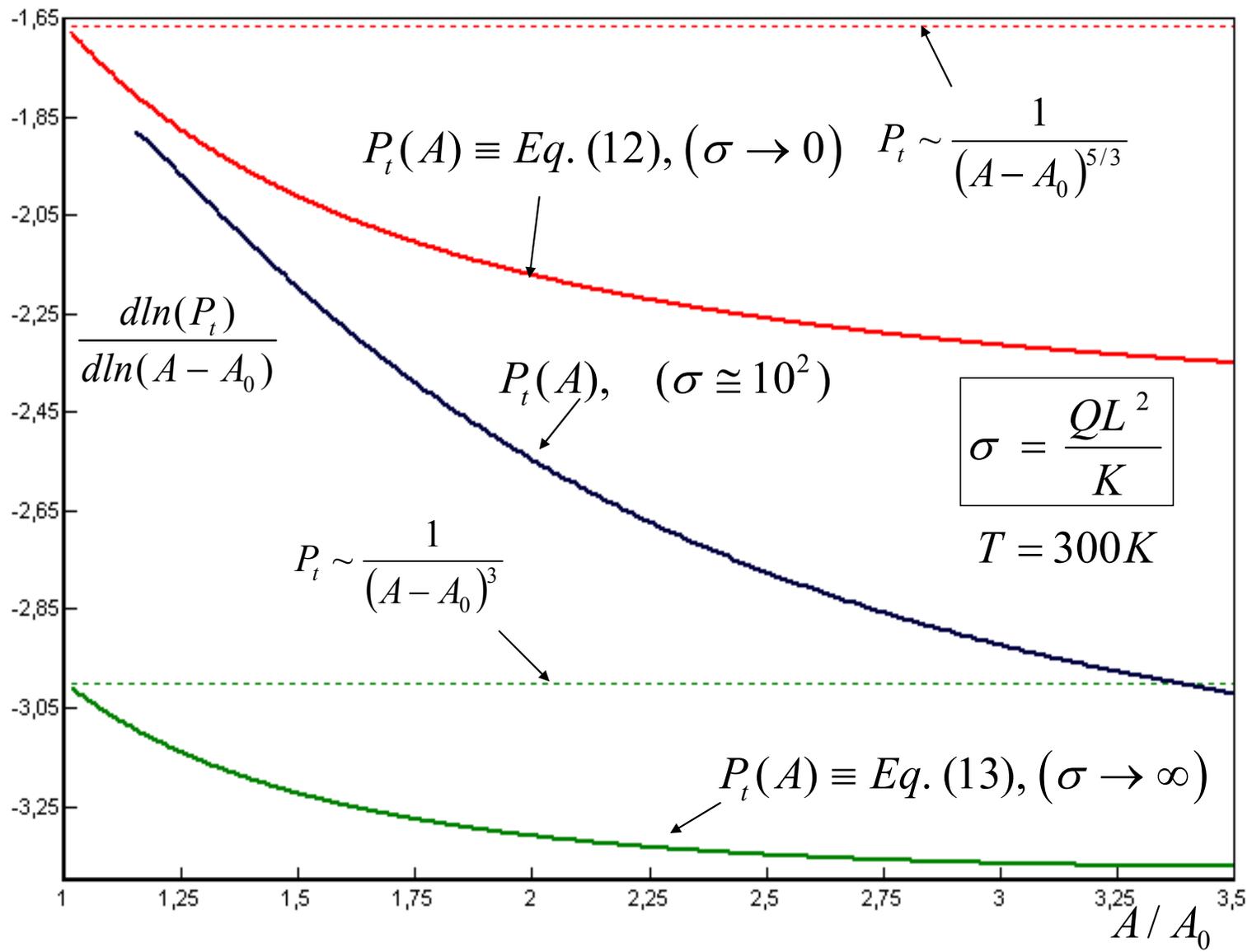

Fig. 3

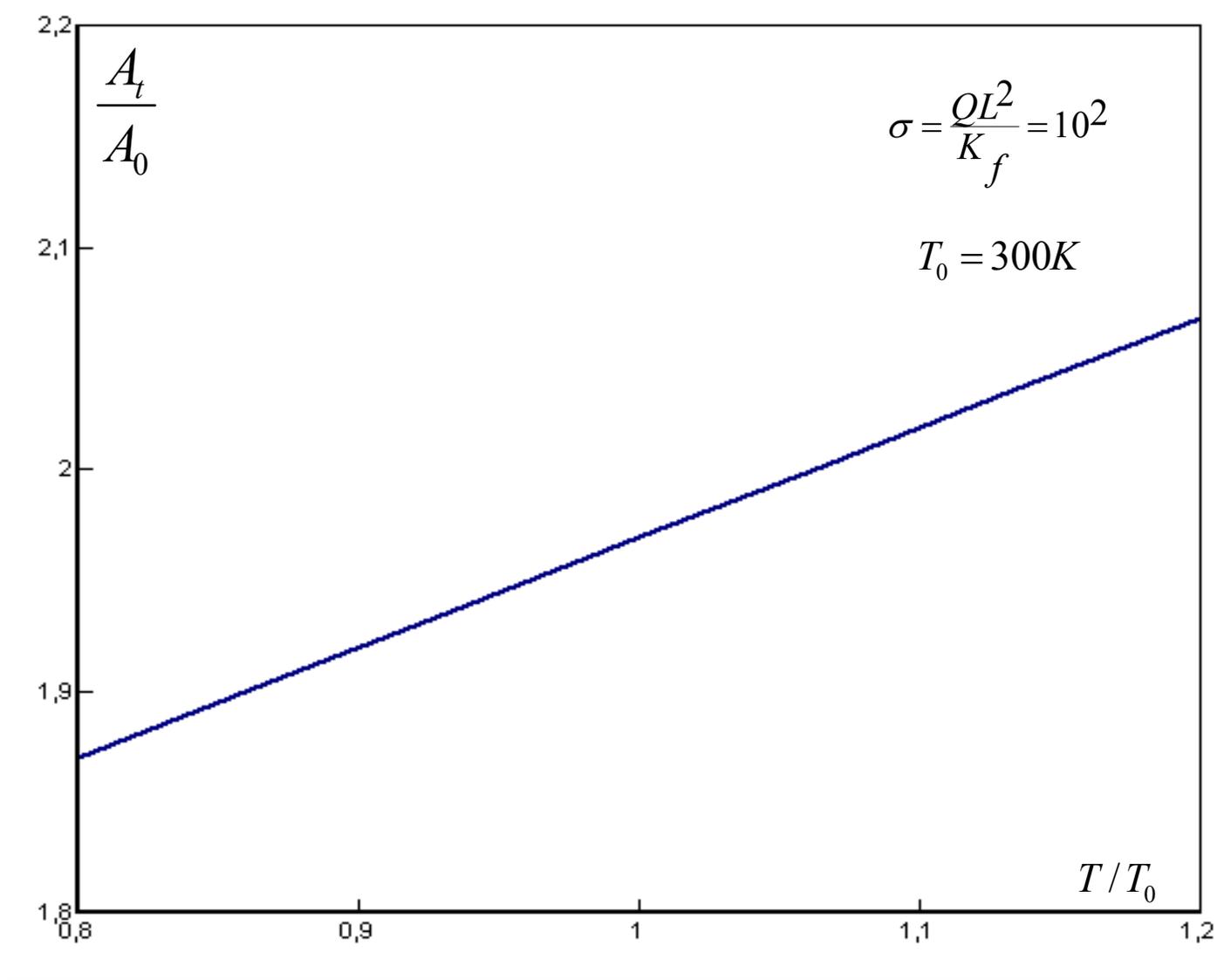

Fig. 4

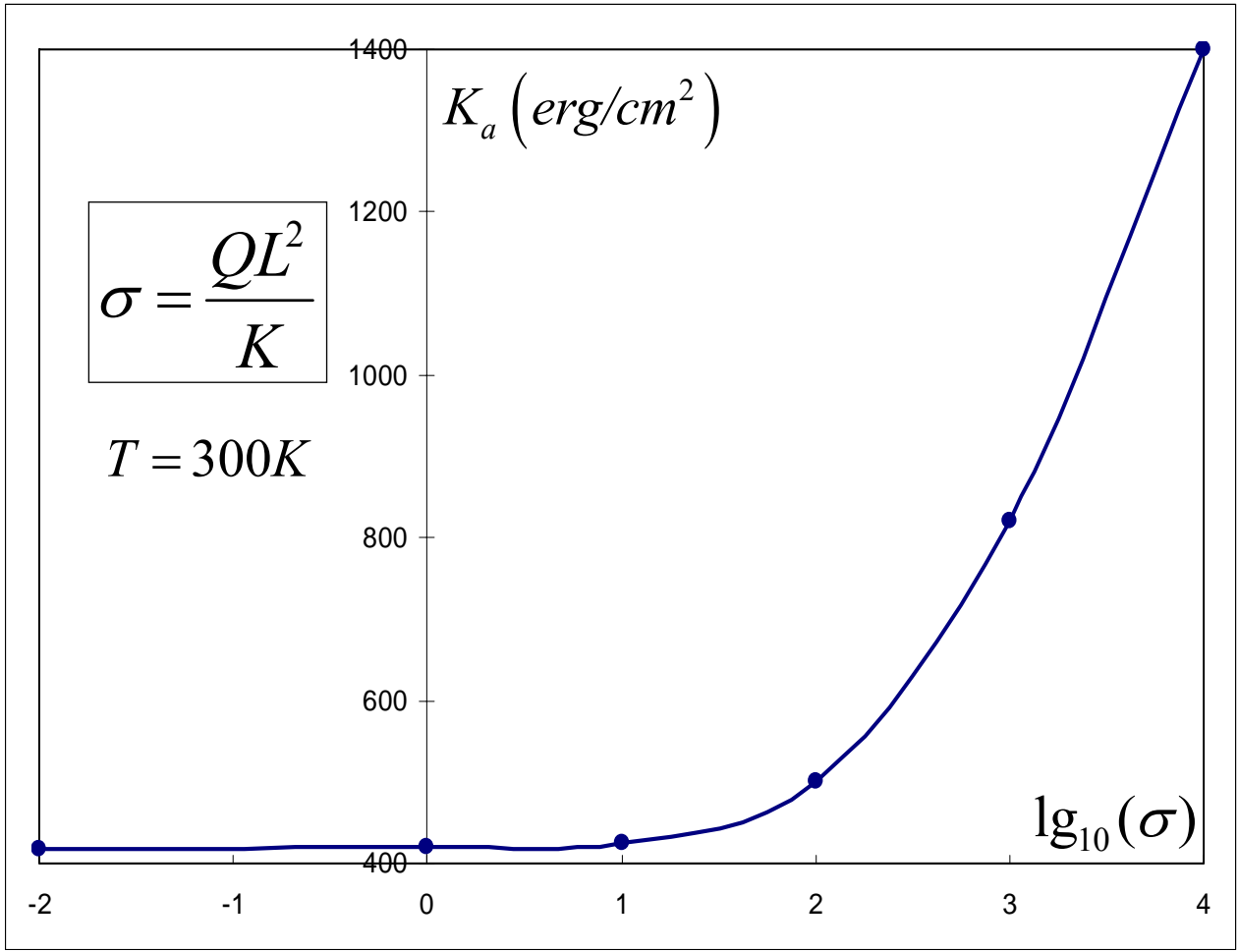

Fig. 5

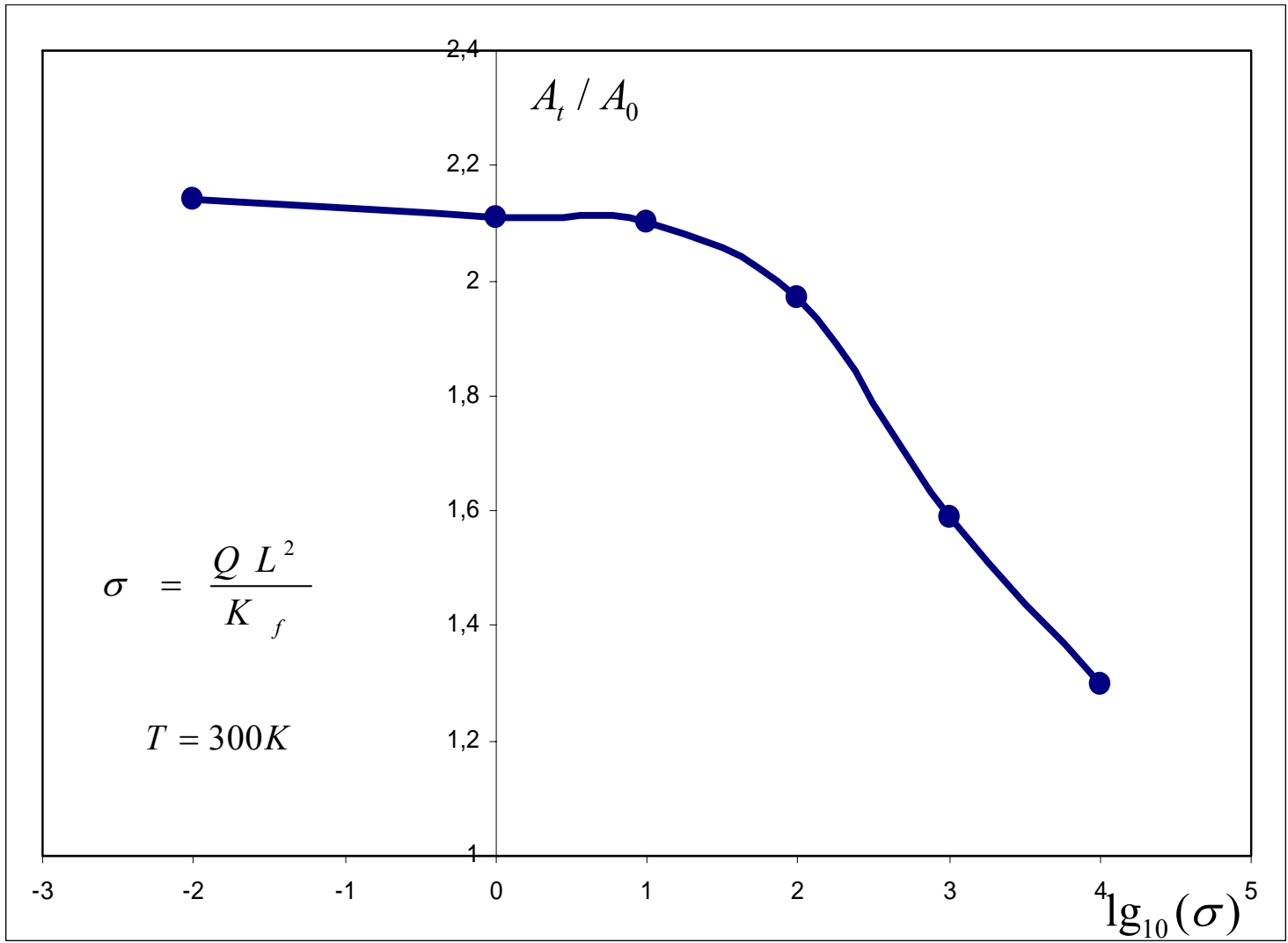

Fig. 6

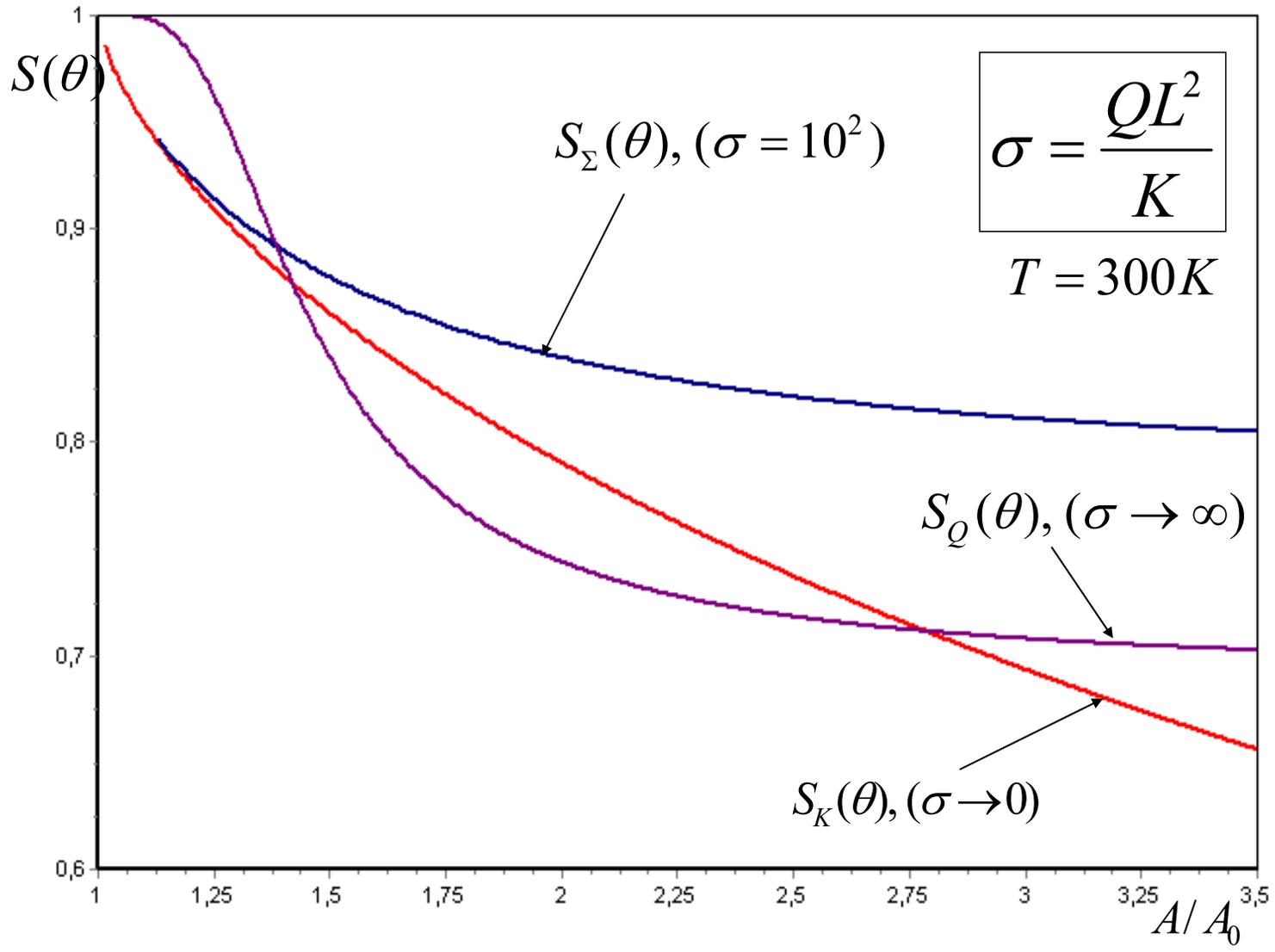

Fig. 7